\documentclass[a4paper,11pt]{article}
\pdfoutput=1 
\usepackage[utf8]{inputenc}  % added by arXiv 

\usepackage{jinstpub}
\usepackage{graphicx} 
\usepackage{url}
\usepackage{multirow}
\usepackage{amsmath}

\usepackage[dvipsnames]{xcolor}

\title{\boldmath Front-end control system and precise threshold configuration of the \texorpdfstring{$\nu$}~-Angra experiment}

\author[a]{Mariana L Migliorini,}
\author[a]{Antônio Fernandes Jr,}
\author[b]{João C Anjos,}
\author[c]{Pietro Chimenti,}
\author[a]{Igor A Costa,}
\author[d]{Luis F G Gonzalez,}
\author[e]{Germano P Guedes,}
\author[d]{Ernesto Kemp,}
\author[b]{Herman P Lima Jr,}
\author[a]{Guilherme S P Lopes,}
\author[a]{Amaro S Lopes Jr,}
\author[a]{Rafael A Nobrega\footnote{Corresponding author.},}
\author[a]{Igor F Pains,}
\author[f]{Iuri M Pepe,}
\author[f]{Dion B S Ribeiro,}
\author[a]{David M Souza}

\affiliation[a]{Universidade Federal de Juiz de Fora, Juiz de Fora, Brazil}
\affiliation[b]{Centro Brasileito de Pesquisas Físicas, Rio de Janeiro, Brazil}
\affiliation[c]{Universidade Estadual de Londrina, Londrina, Brazil}
\affiliation[d]{Universidade Estadual de Campinas, Campinas, Brazil}
\affiliation[e]{Universidade Estadual de Feira de Santana, Feira de Santana, Brazil}
\affiliation[f]{Universidade Federal da Bahia, Salvador, Brazil} 

\emailAdd{rafael.nobrega@ufjf.edu.br}

\date{March 2020}

\abstract{
The $\nu$-Angra experiment aims to estimate the flux of antineutrino particles coming out from the Angra II nuclear reactor. Such flux is proportional to the thermal power released in the fission process and therefore can be used to infer the quantity of fuel that has been burned during a certain period. To do so, the $\nu$-Angra Collaboration has developed an antineutrino detector and a complete acquisition system to readout and store the signals generated by its sensors. The entire detection system has been installed inside a container laboratory placed beside the dome of the nuclear reactor, in a restricted zone of the Angra II site. The system is supposed to work standalone for a few years in order to collect enough data so that the experiment can be validated. The detector's readout electronics and its environmental conditions are crucial parts of the experiment and they should work autonomously and be controlled and monitored remotely.
Additionally, threshold configuration is a central issue of the experiment since antineutrino particles produce low energy signals in the detector, being necessary to carefully adjust it for all the detector channels in order to make the system capable of detecting signals as low as those generated by single photons.
To this end, an embedded system was developed and integrated to the detection apparatus installed in the container at the Angra II site and is now operational and accessible to the $\nu$-Angra Collaboration.
This article aims at describing the proposed embedded system and presenting the results obtained during its commissioning phase.}

\begin{document}
\maketitle
\flushbottom

\section*{Introduction}

The use of embedded systems for remote control and monitoring of experiments has been increasing in several areas of research. Its application facilitates management and reduces cost of scientific experiments, allowing measurement conditions to be controlled and monitored even over great distances. In the context of experimental particle physics, this need is even more present given that in many cases the access to the experiment site is restricted.
The Neutrinos Angra ($\nu$-Angra) experiment \cite{anjos2012status,anjos2015using,comiss1,comiss} is part of a global nuclear-fuel safeguard effort \cite{bowden2008reactor}, coordinated by the IAEA (International Atomic Energy Agency). Antineutrino detectors can be used to measure the thermal power of reactors and are sensitive to the burn-up process of nuclear fuel.
The $\nu$-Angra experiment aims to develop a surface antineutrino detector to be used as a tool for monitoring the fuel burn-up process in nuclear reactors, making it possible to infer the amount of nuclear fuel used in each energy production cycle.
The detector is currently in operation next to the dome of the Angra II reactor in Angra dos Reis, Brazil, and tests and data analysis are in progress. The detection system was mounted inside a container in a restricted area, 25 m away from the reactor core.

\section{The \texorpdfstring{$\nu$}~-Angra measurement system}

\begin{figure}[h!]
	\centering
	\includegraphics[width=0.7\linewidth]{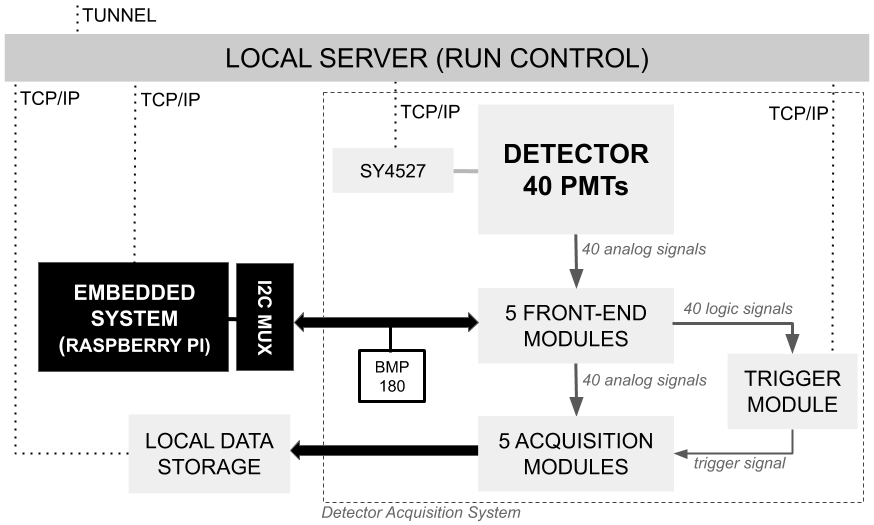}
  	\caption{Schematic diagram of the $\nu$-Angra system mounted few meters away from the Angra II nuclear reactor dome.}
  	\label{fig1}
\end{figure}

The measurement system installed in the container was designed to work autonomously.
A local server provides connection to the external network through IP tunneling and has access to all the subsystems that make up the system.
A program named \textit{Run Control} running in this local server is responsible for organizing and executing data acquisition runs through the control and monitoring of the whole system. This program uses the TCP/IP protocol as a means of exchanging messages and commands with the experiment subsystems.
Figure~\ref{fig1} shows a schematic diagram of this system.

\clearpage

In addition to the local server and the detector itself, the measurement system is composed of the following subsystems:
sensor readout electronics (front-end modules) \cite{dornelas2016front}, acquisition and trigger electronics \cite{lima2014data,trigger}, data storage, high-voltage system from CAEN model SY4527 \cite{hvcaen}, a device with pressure and temperature sensors (BMP180) \cite{sensortec2013data} and an embedded system.
This embedded system, highlighted in figure~\ref{fig1} (in black), and its threshold configuration procedure are the central subject of this article.
Before going into details about them, it is important to describe the general characteristics of the readout and acquisition electronics of the experiment.

The detector is filled with water using the Cherenkov effect as the main process for particle detection. It can be divided into two subsystems: an inner detector called TARGET that defines a fiducial volume for the search of antineutrino events, and an outer detection system called VETO, used to identify events produced by natural radioactivity particles. More details are given in Ref. \cite{anjos2012status, anjos2015using}.
The whole detector is instrumented with 40 Photomultiplier Tubes (PMTs) model R5912 by Hamamatsu \cite{hamamatsu1998photomultiplier, Zhang2016}, 32 installed in the TARGET detector and 8 in the VETO, operating at a gain of $1 \times 10^7$ electrons per photoelectron. Each PMT generates impulse signals that are sent to the front-end modules, which are composed of Amplifier-Shaper-Discriminator (ASD) circuits.
The output of the amplification/shaping stage is subdivided into two branches, one that is sent to the acquisition system where signals are digitized and stored in-board to wait for a trigger decision, and another that is sent to the input of a discriminator, also part of the front-end board, where the impulse signal is compared to a voltage level (threshold) to generate logic pulses which are finally sent to the trigger system.
The two adjustable parameters of a front-end channel are its discriminator threshold and the pedestal (or offset) of the signal which is sent to the acquisition system.
Each front-end module contains eight independent channels for reading out PMT signals, and each channel has its own threshold and pedestal levels that can be configured independently of each other.
To control all these pedestal and threshold values individually, each front-end module contains four AD5645R Integrated Circuits (ICs) \cite{ad5665r200212} connected to a single $I^2C$ bus.
Each AD5645R CI provides four Digital-to-Analog Converter (DAC) channels of 14 bits.
The embedded system uses custom routines to adjust and monitor all the 80 DAC channels and to monitor atmospheric pressure and temperature of the experiment. These two tasks are motivated below.

\paragraph{Atmospheric pressure and temperature:} the internal volume of the container should be maintained at a temperature below 25 $^oC$ for reasons of stability and protection of the components that are part of the system.
In addition, atmospheric pressure measurements can also help in understanding the detector's acquired data since the rate and energy of background events caused by cosmic rays \cite{PT} are influenced by it.

\paragraph{Discriminator threshold and signal pedestal:} threshold adjustment is of paramount importance due to the fact that antineutrino events are in a low energy region, which may generate impulse signals of low amplitude in some channels of the detector \cite{anjos2015using}.
In order to detect antineutrinos with good efficiency, the threshold of each channel must be adjusted to a level slightly above the electronic noise signal.
Given its importance, an accurate and precise threshold configuration procedure has been proposed and implemented in the context of this work.
Pedestal adjustment is used to give a greater dynamic range for the signals produced by the front-end circuit since they saturate whenever an amplitude greater than $1.4~V$ is reached at the output of the front-end amplification/shaping stage~\cite{dornelas2016front}. A negative pedestal should therefore be used for this purpose.
Finally, threshold and pedestal values should be continually monitored to ensure that the established settings remain valid after weeks/months of detector operation.

%%%%%%%%%%%%%%%%%%%%%%%%%%%%%%%%
%%%%%%%%%%%%%%%%%%%%%%%%%%%%%%%%
%%%%%%%%%%%%%%%%%%%%%%%%%%%%%%%%
\section{Front-end control system} \label{sec:develop}

In this section, the modules that make up the embedded system designed to control and monitor important parameters of the $\nu$ -Angra experiment will be described.

\subsection{Hardware and application layer}
\begin{figure}[ht]
	\centering
	\includegraphics[width=0.35\linewidth]{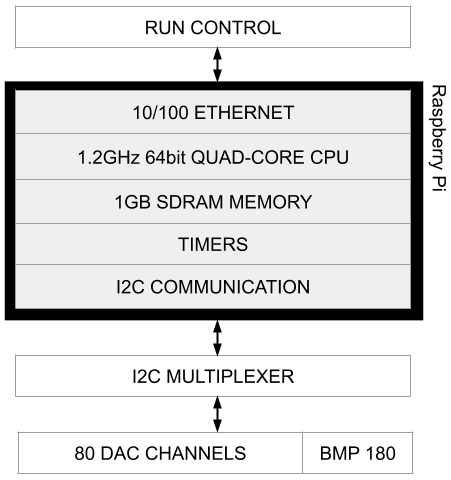} ~~~~~
	\includegraphics[width=0.35\linewidth]{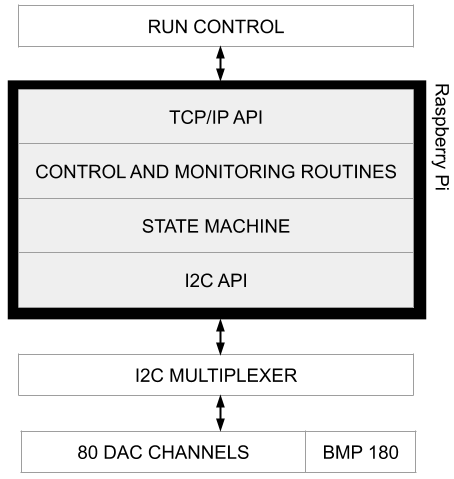}
	\caption{Hardware (left) and application layer (right) schematics of the embedded system.}
\label{fig3}
\end{figure}

The front-end control system makes use of a low cost embedded system composed of a Single Board Computer (SBC) based on a \textit{Raspberry Pi 3 Model B}.
Such a system is accessed by Ethernet and is connected to six front-end modules and to a BMP180 device via $I^2C$ protocol.
The BMP180 is composed of a piezo-resistive sensor, an ADC, a control unit with E$^2$PROM and an $I^2C$ serial interface.
A mezzanine board based on the TCA9548A IC \cite{TCA9548A2016} (\textit{I2C MUX} in figure~\ref{fig1}) offers an $I^2C$ bi-directional switch supporting eight channels which might operate in \textit{Standard} mode (100~kbit/s) or \textit{Fast} mode (400 kbit/s).
Such board provides several $I^2C$ channels which are required to carry out the communication between the SBC and the external devices.
Figure~\ref{fig3} on the left shows a schematic of this system.

\begin{figure}[ht]
	\centering
	\includegraphics[width=0.39\linewidth]{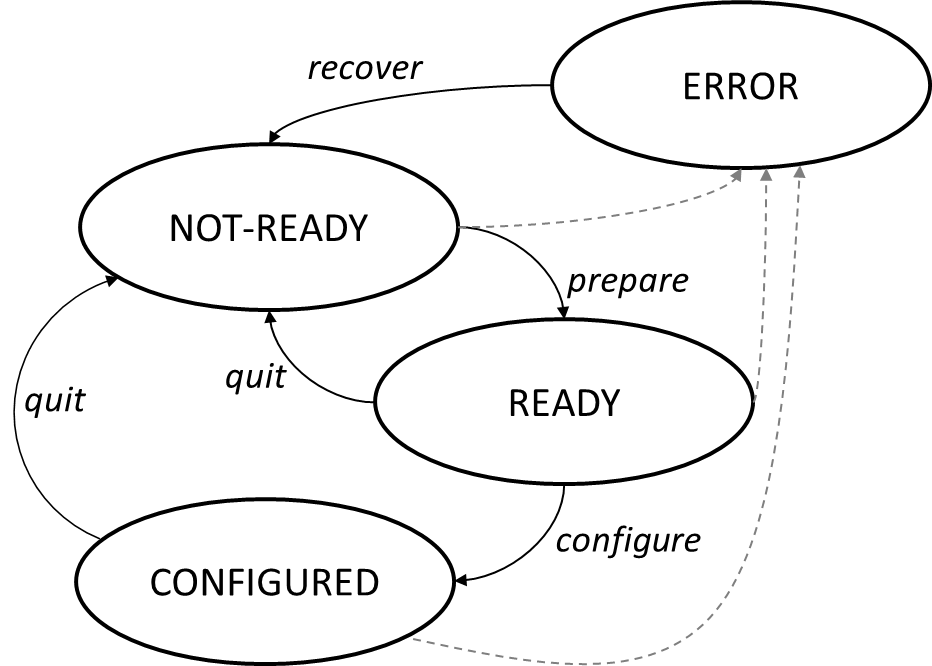}~~~~~~~~~~
	\includegraphics[width=0.5\linewidth]{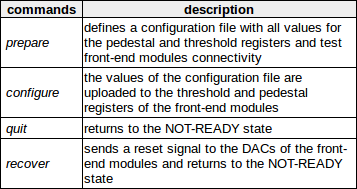}
	\caption{State machine for data collection management and description of its commands.}
\label{fig5}
\end{figure}

The application layer implemented in the embedded system can be divided into four groups as shown on the right of figure~\ref{fig3}: (1)~definition and management of TCP/IP messages (\textit{TCP/IP~API}); (2)~control and monitoring routines; (3)~state machine; and (4) definition of $I^2C$ commands (\textit{I2C~API}).
A TCP/IP interface was implemented locally in the SBC by means of a \textit{Python} script which uses the fundamentals of the \textit{Socket} API \cite{python}. The $I^2C$ interface was implemented using the \textit{smbus} module for \textit{Python} allowing access to the system's $I^2C$ devices.
Finally, a state machine, as represented in figure~\ref{fig5}, was implemented to be used by the \textit{Run~Control} in the organization and execution of data acquisition runs.

\subsection{Control and monitoring routines}
\label{sec:routines}

The main routines implemented by the control system are: (1) temperature and pressure measurements; (2) configuration and monitoring of signal pedestal and discriminator threshold values; and (3) threshold scan procedure. They are described below.

\paragraph{Atmospheric pressure and temperature measurements:}
Pressure and temperature measurements are stored in a local database, in the embedded system. Each measurement is accompanied by date and time in a format known as \textit{Unix~Timestamp}. The \textit{Run~Control} can configure the measurement rate and access this database via TCP/IP whenever needed.

\paragraph{Pedestal and threshold configuration and monitoring:}
different values of pedestal and threshold may be necessary to create a greater understanding of the operational characteristics of the detector and of the collected events. To make it possible, different configuration files, containing all the values of pedestal and threshold, were created (e.g. \textit{PHYSICS1}, \textit{PHYSICS2} and \textit{TEST}).
According to the message received via TCP/IP, the values specified in one of these files are uploaded into the front-end registers to set pedestal and threshold values.
From this moment on, register values can be monitored with a rate that can be defined by the \textit{Run Control}.

\begin{figure}[ht]
	\centering
	\includegraphics[width=0.85\linewidth]{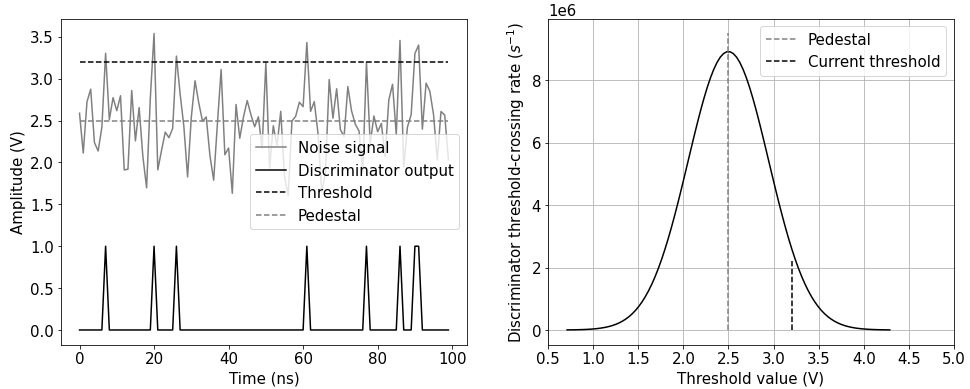}
	\caption{Representation of the threshold scan procedure and its resulting curve at right.}
\label{fig4}
\end{figure}

\paragraph{Threshold scan procedure:}
this routine was implemented to allow precise adjustment of threshold and noise monitoring for each detector channel. In this routine, the threshold is swept downwards from an initial value well above the pedestal level and for each new threshold value, the output signal rate of the front-end discriminator is measured to form a signal rate vs threshold curve as illustrated in figure~\ref{fig4}. It should be taken into account that, for this routine, the trigger system measures the number of times the noise signal crossed the threshold level, considering only the signal rising edge. The threshold scan produced curve (on the right side of figure~\ref{fig4}) tends to be bell shaped whose characteristics depends on the bandwidth of the noise signal and on the discriminator timing response \cite{Anderlini_2013}.

In this procedure, therefore, the threshold moves from top to bottom and as it approaches the signal pedestal, the discriminator signal rate increases due to the presence of noise. The peak value of the resulting curve can then be used to calibrate the channel pedestal relating it to a corresponding threshold value. When this routine is applied with the detector turned on, it is expected to see other components besides electronic noise due to the interaction of particles in the detector and the occurrence of dark current pulses \cite{baicker1960dark}. Section \ref{sec:trscan} will present a study of this technique in order to understand better its characteristics and possible applications.

\subsection{Laboratory setup for development and testing}
\begin{figure}[ht]
	\centering
	\includegraphics[width=0.7\linewidth]{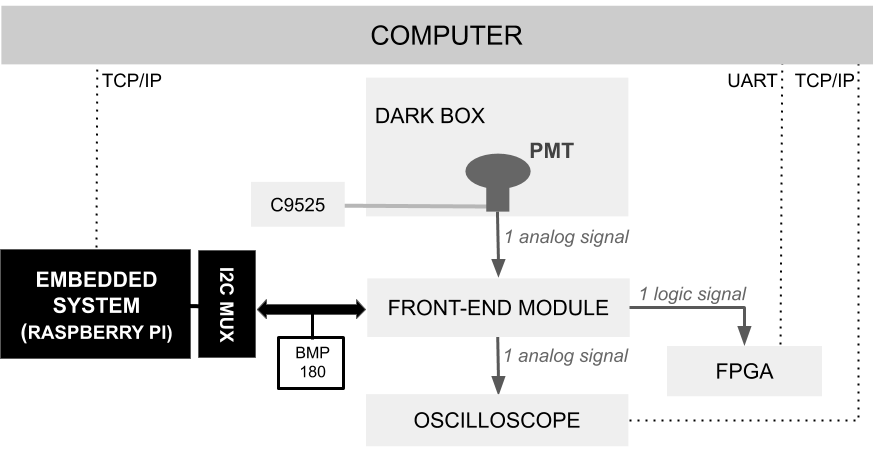}
  	\caption{Schematic diagram of the setup assembled in laboratory for development and validation of the control system and its routines.}
  	\label{fig6}
\end{figure}

A test setup was assembled in laboratory to serve as a basis for the development and testing of the embedded system and its routines.
Figure~\ref{fig6} shows a schematic diagram of this setup that includes, in addition to the embedded system, a computer, a dark box with a PMT (same R5912 model used in the $\nu$-Angra detector) powered by a Hamamatsu C9525 power supply \cite{hamamatsuC9525}, a front-end module, an FPGA module and an oscilloscope, besides the sensors provided by the BMP180 circuit.

The signals generated by the PMT are sent to the front-end module, which in turn sends its analog output signals to the oscilloscope and its logic signals to the FPGA module.
The latter communicates with the computer using the UART protocol through an USB-TTL converter circuit while the oscilloscope is controlled via TCP/IP.
The FPGA module emulates the Trigger electronics of the experiment while the oscilloscope replaces the acquisition module.
The embedded system controls the discriminator threshold and signal pedestal values of the front-end module and measures the atmospheric pressure and ambient temperature through the BMP180 device via $I^2C$.
Therefore, this setup emulates part of the measurement system installed at the Angra II nuclear power plant site, making it possible to develop the embedded system routines and test them before implementation in the experiment.

Finally, the single photoelectron (SPE) spectrum has been measured based on the front-end output signal in terms of peak amplitude in mV units. The high-voltage of the PMT sensor was set to 1510~V, which was the value provided by its manufacturer to make it to work with a gain of $1 \times 10^7$, and a light-emitting diode was placed in front of the PMT cathode and properly pulsed by a waveform generator. 
The acquired data were fitted by a function made up of a sum between an exponential and a Gaussian function.
The former models the noise component, while the latter models the SPE spectrum.
The mean value and standard deviation of the SPE spectrum were estimated at 79.5 $\pm$ 0.7 mV and 21.7 $\pm$ 0.6 mV, respectively.

\begin{figure}[ht]
\centering
\includegraphics[width=0.7\linewidth]{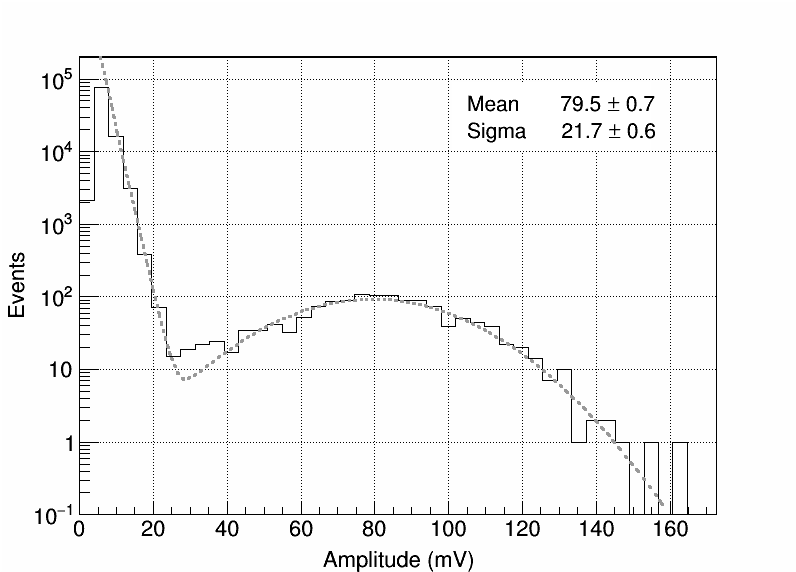}
\caption{SPE spectrum measured at the front-end output.}
\label{spe_spec}
\end{figure}

\section{Results}
\label{sec:results}

This section is divided into three topics: (1) temperature and atmospheric pressure test measurements, (2) study of the threshold scan procedure performed in laboratory, and (3) its application to calibrate and configure the detector channels at the nuclear reactor site.

\subsection{Atmospheric pressure and temperature measurements}

\begin{figure}[ht]
\centering
\includegraphics[width=0.50\linewidth]{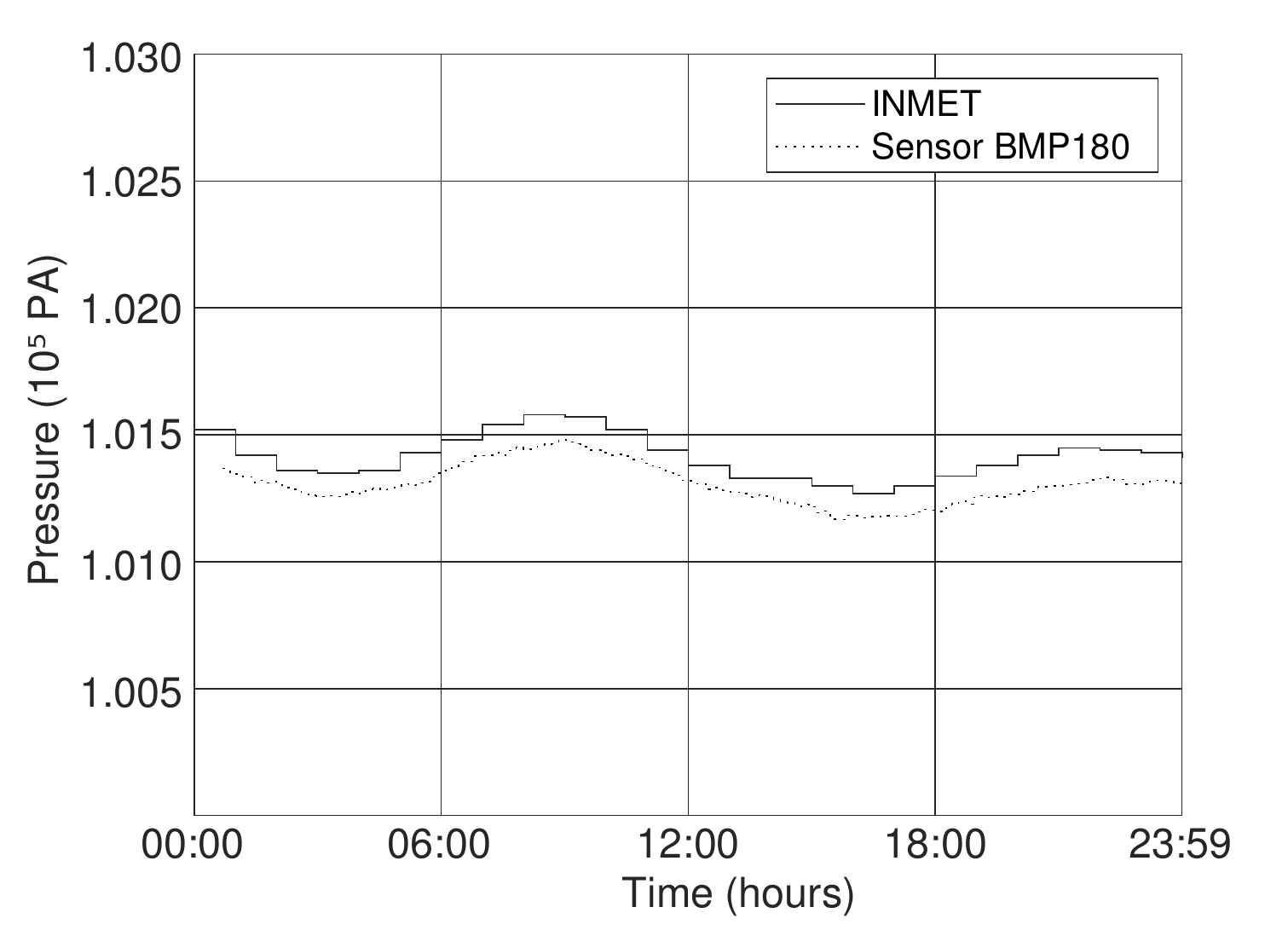}
\includegraphics[width=0.475\linewidth]{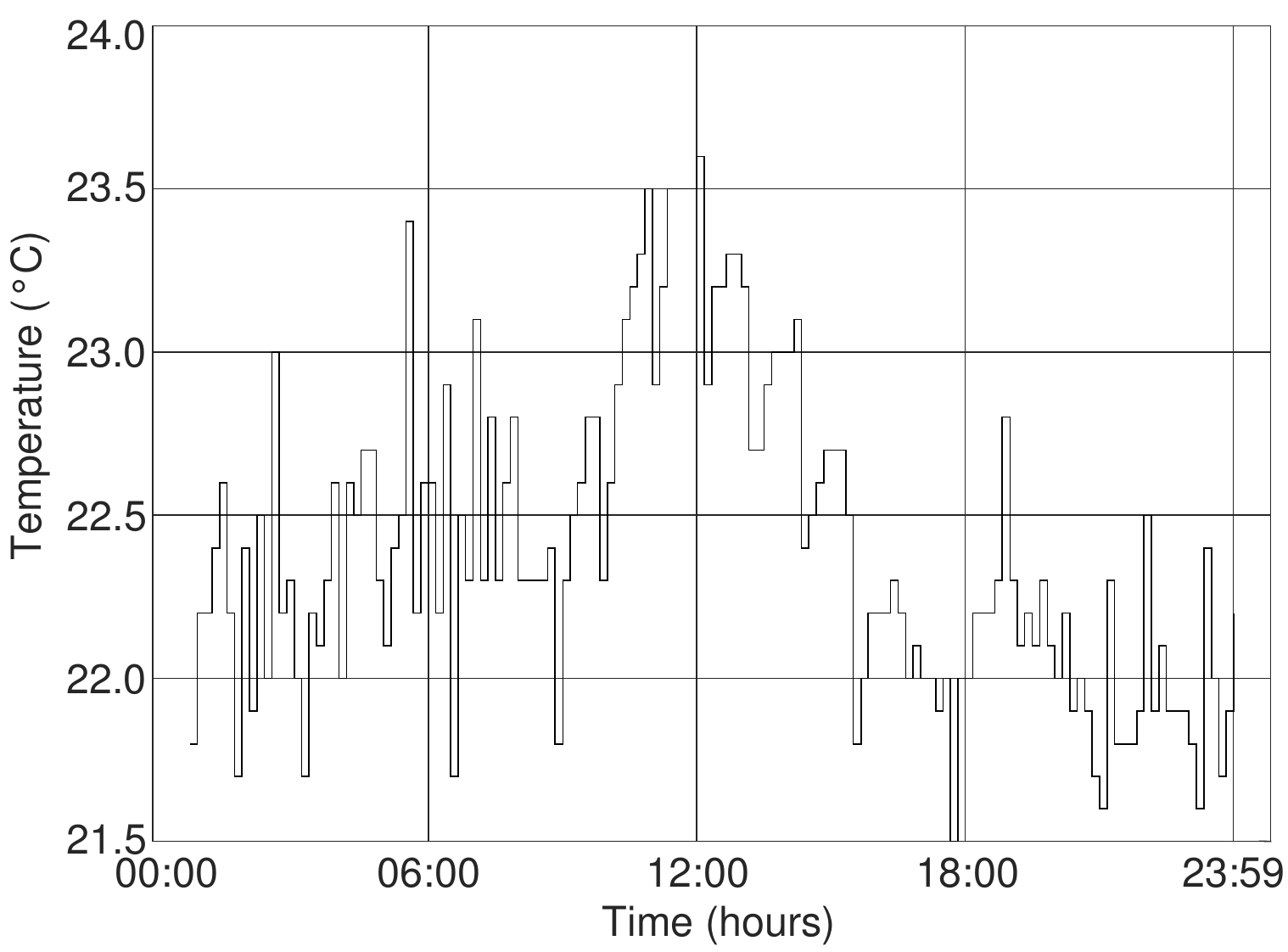}
\caption{Atmospheric pressure and temperature measurements considering a full day of detector operation.}
\label{temp}
\end{figure}

In one day, the local atmospheric pressure fluctuates mainly due to the temperature variation and the force of gravity exerted by the moon.
Pressure fluctuations occur throughout the day, peaking around 10AM and decreasing around 4PM.
Figure~\ref{temp} (left) shows that the installed pressure sensor has sufficient resolution to detect this oscillation.
This measurement was performed next to the dome of the Angra II plant, resulting in an average pressure of approximately $ 1.013 \times 10^5~Pa $.
This measurement was compared with data collected in the region of Angra dos Reis by the National Meteorological Institute (INMET) \cite {INMET}. It can be seen from figure~\ref{temp} on the left that the measurements have the same oscillation period, as expected, with a very small difference between them, which can be justified by the different locations where the sensors were installed.
The same measurement was performed in the city of Juiz de Fora, Minas Gerais, at an average altitude of 678~m, resulting in an average atmospheric pressure of $0.918 \times 10^5~Pa $.
According to INMET data, in Juiz de Fora the atmospheric pressure usually varies from approximately $ 0.905 \times 10^5~Pa $ (summer) and $ 0.923 \times 10^5~Pa $ (winter).

The temperature sensor was installed next to the detector's readout electronics, inside the container and therefore under the influence of air cooling. Figure~\ref{temp} (right) shows the related measurement for a full day of detector operation. 
As can be seen, the temperature peaked around noon, reaching a value of 23.6 $^oC$.
Note that the high frequency oscillations of the curve occur due to the air conditioning on-and-off process. These measures will allow the ambient temperature of the laboratory to be monitored continuously during the operation of the experiment.

\subsection{Evaluation of the threshold configuration procedure in laboratory}
\label{sec:trscan}

In this section the threshold scan procedure will be studied in different PMT conditions; with it turned-off to generate signal with electronic noise only, and turned-on to check for other signal components.
For testing purposes, the pedestal level was set to $300~mV$ and the threshold value swept in decreasing mode. The initial threshold value was programmed to be well above the pedestal value, that is, far from the noise region. For each new threshold value, the event rate at the output of the front-end discriminator was measured by the FPGA module (see figure~\ref{fig6}), forming a graph of signal rate vs threshold as described previously in section \ref{sec:routines}.

\begin{figure}[ht]
\centering
\includegraphics[width=0.49\linewidth]{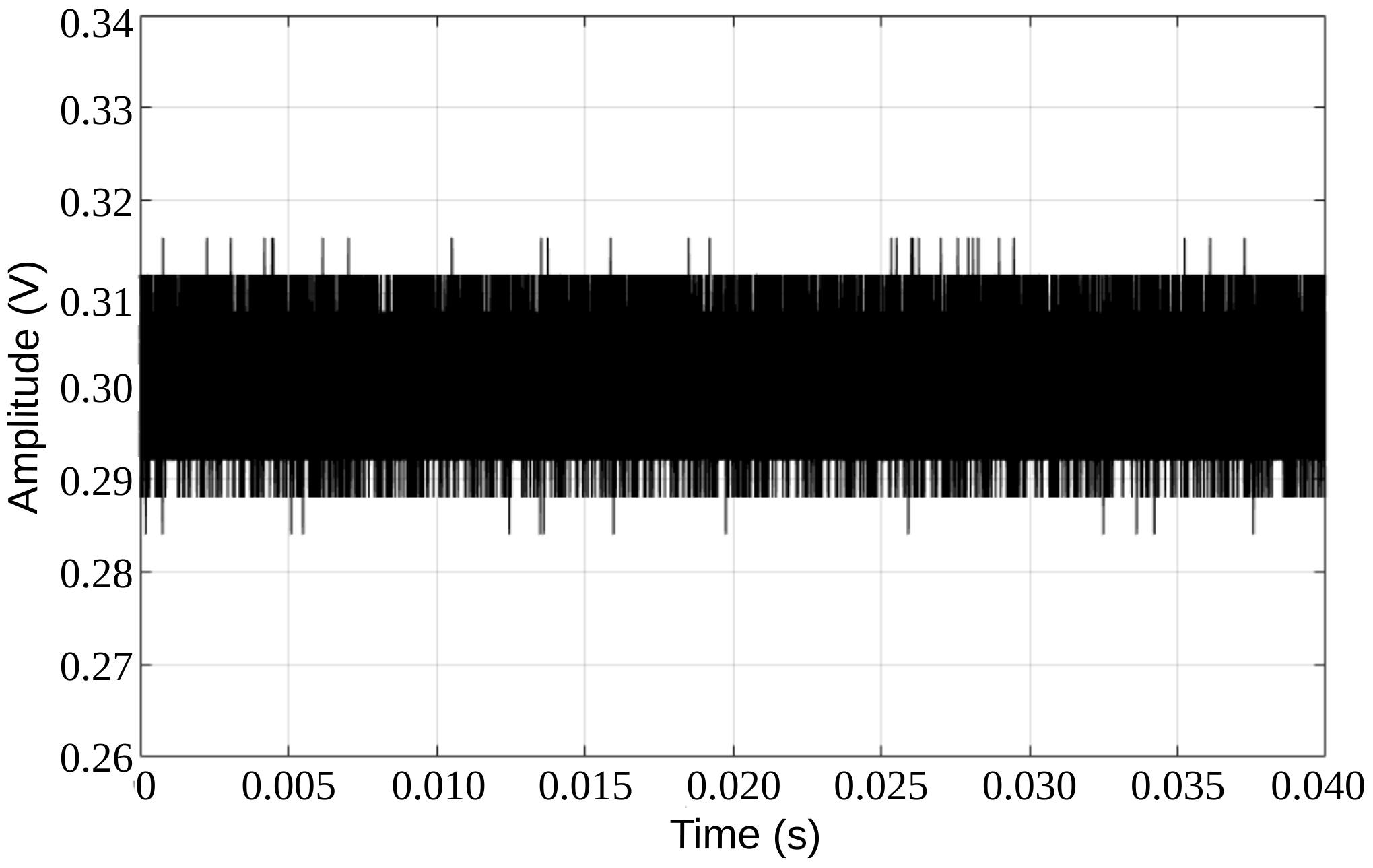}
\includegraphics[width=0.49\linewidth]{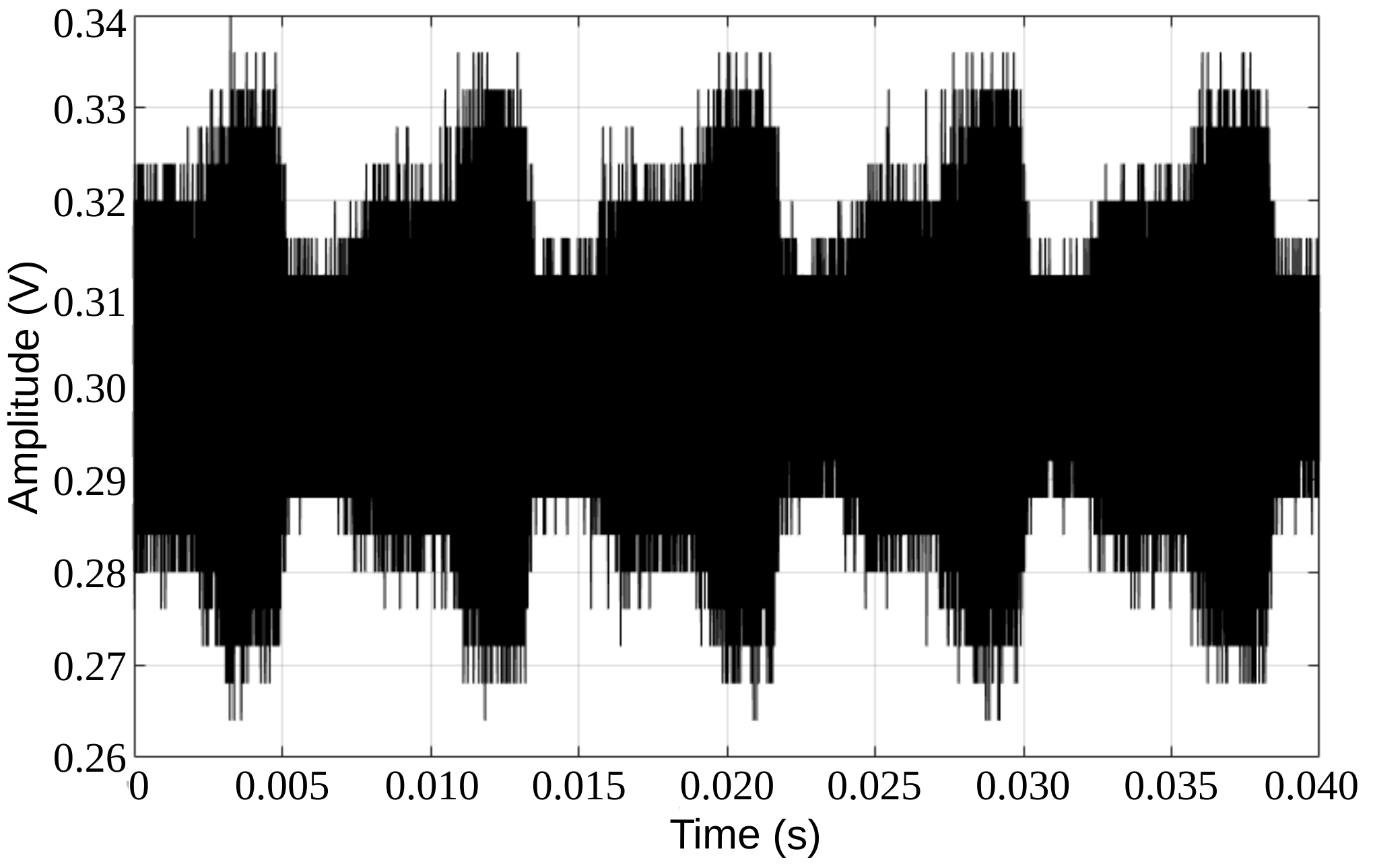}
\caption{Noise at the output of the front-end channel when its input is terminated with $ 50 \Omega $ (left) and when its input is connected to the PMT signal cable but with its high-voltage power cable floating (right).}
\label{noise_fe}
\end{figure}

\begin{figure}[!h]
\centering
\includegraphics[width=0.57\linewidth]{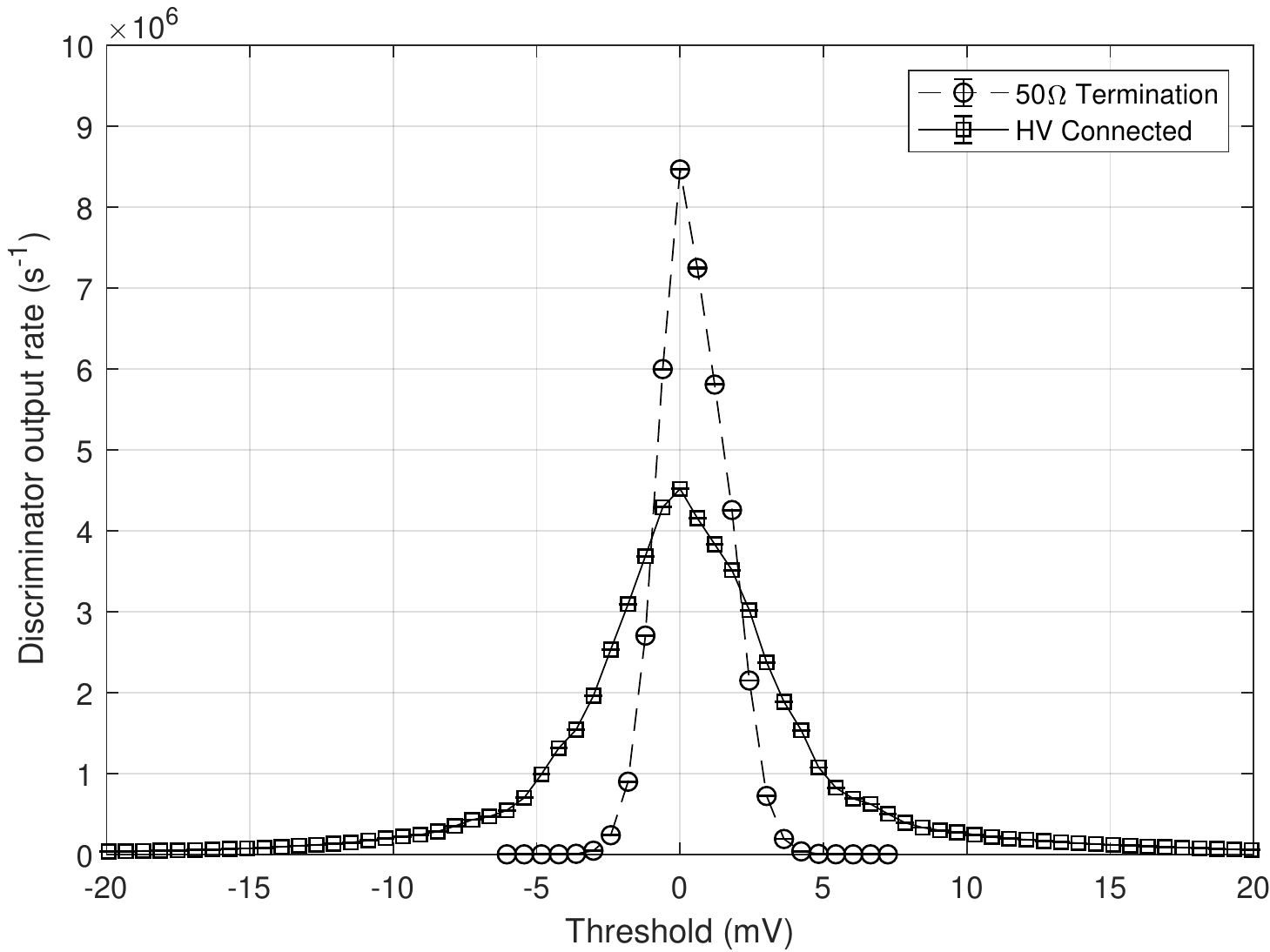}
\caption{Threshold scan curves for the two noise situations shown in figure \ref{noise_fe}. These curves were centered at zero from their peak values.}
\label{NoInput_Ch1}
\end{figure}

For a first test, two noise situations were created. One with the front-end channel input connected to a $50 \Omega$ termination (no use of PMT) and another with the PMT signal cable connected to the front-end input but with the PMT power cable floating. In this last configuration, the high-voltage cable undergoes electromagnetic interference increasing the noise level arriving to the front-end circuit. Figure~\ref{noise_fe} shows the signals at the output of the front-end for both cases. The noise standard deviation were measured to be 3.5 mV and 5.8 mV, respectively.
Figure~\ref{NoInput_Ch1} shows the resulting threshold scan curve for those cases. This curve can be used to calibrate the threshold in relation to the signal pedestal and to measure and monitor noise throughout the experiment data acquisition period.

\begin{figure}[ht]
\centering
\includegraphics[width=0.7\linewidth]{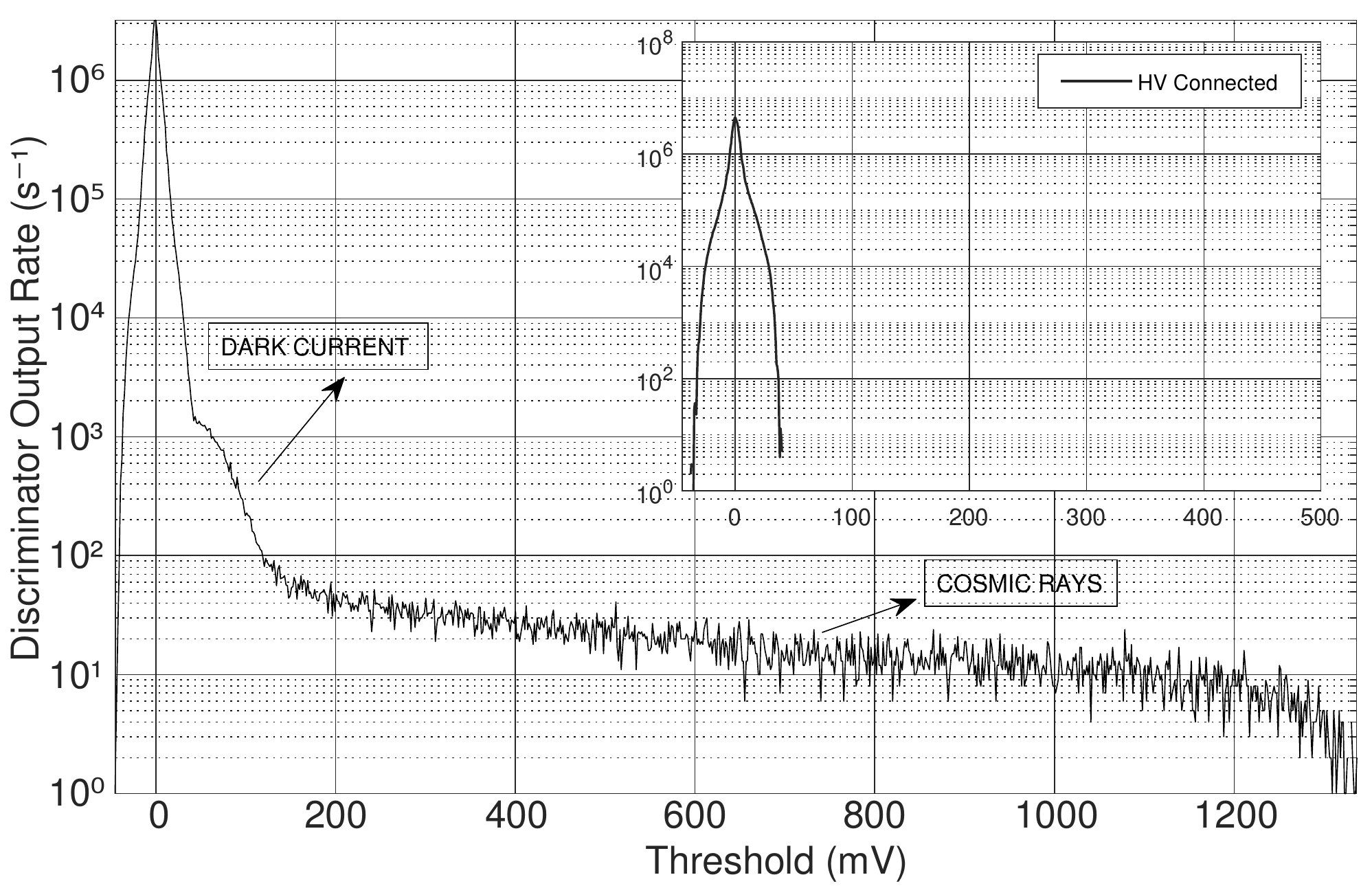}
\caption{Threshold scan curve obtained with a PMT powered with 1510 V and connected to the front-end input. The internal graph corresponds to one of the measurements presented in figure~\ref{NoInput_Ch1}, but with the horizontal and vertical axes changed to allow a direct comparison with the curve shown in the main frame.}
\label{cnt_HVon}
\end{figure}

In a different scenario, used to understand the threshold scan procedure in a situation where the PMT is turned-on, the PMT power cable was finally connected to the high-voltage power supply which was configured to provide a voltage of $ 1510~V $ setting the PMT gain to $1 \times 10 ^ 7 $ electrons per photoelectron, as used in the experiment. Figure~\ref{cnt_HVon} shows the resulting threshold scan curve.
In addition to the electronic noise component, shown in figure~\ref{NoInput_Ch1}, it is possible to identify two other components: one in the region between $ 40~mV $ and $ 120~mV $ and another after $ 120~mV $.
The former occurs due to dark current events \cite{baicker1960dark}.
Its causes include thermionic emission, field effects and leakage currents. The characteristics of this process depend particularly on the composition of the cathode and, throughout the usual range of supply voltage, the average amplitude of the signals generated by this effect varies proportionally with the gain of the PMT \cite{DCganho}.
Therefore, the identification of the dark current events could be used to monitor the gain variation of the detector PMTs.
The latter is related to cosmic ray events that generate photons through Cherenkov effect \cite{mostafa2014high}, producing signals with a wide range of amplitude values.

The main error sources of the threshold scan measurement are the DAC \textit{integral nonlinearity error} which can reach, in the worst case, 4$\times$LSB or, equivalently, 1.2~mV for the AD5645R chip \cite{ad5665r200212}, and its \textit{gain error} which can get to a value of -0.05\% of the Full Scale Range (FSR) for a temperature of 100~$^o$C, also considering the worst case. The front-end electronics uses a FSR of 5~V for its DACs, causing the \textit{gain error} to reach a value of up to -2.5~mV. At a scale of 100~mV, which represents a value closer to that used for the front-end thresholds, this error reduces to -0.05~mV. The threshold scan procedure cancels the DAC error component known as \textit{offset error} and, therefore, it should not be taken into account. Finally, the uncertainty given by the threshold scan voltage step is approximately $\pm$ 0.3~mV.

The results presented above show the feasibility of using the threshold scan procedure to calibrate and fine tune threshold, and to monitor electronic noise and PMT gain variation of the detector channels.
Such results were obtained using the same PMT model (Hamamatsu R5912) and signal readout electronics as used at Angra II, thus validating the embedded system implemented routines.

\subsection{Threshold calibration and setting at the nuclear reactor site}
\label{sec:realdata}

\begin{figure}[!ht]
\centering
\includegraphics[width=0.49\linewidth]{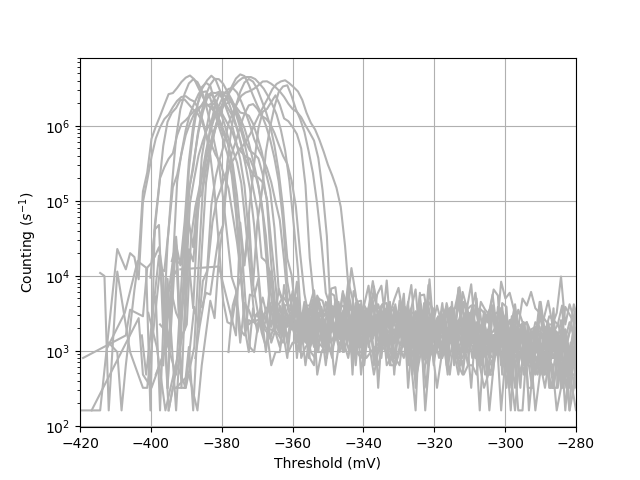}
\includegraphics[width=0.49\linewidth]{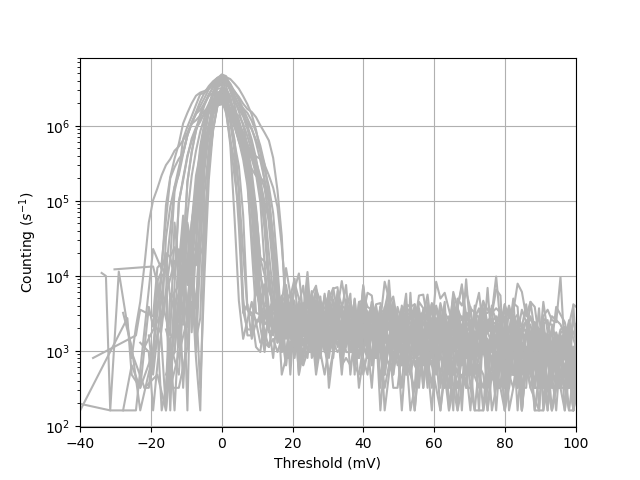}
\caption{Threshold scan curves before (left) and after (right) centering them in relation to their peak values for the TARGET detector channels.}
\label{fig:ThScan1}
\end{figure}

The threshold scan procedure was applied to all the $\nu$-Angra detector channels. The resulting curves are shown in figure~\ref{fig:ThScan1} (left).
Before the threshold scan procedure was available, the experiment thresholds were set assuming that the pedestal values were about the same for all the front-end channels. 
Consequently, a single threshold value of -340 mV was set for all the detector channels in order to ensure that all threshold values were outside the noise region.
However, as shown clearly by the threshold scan curves, a relatively large variation of the effective thresholds (threshold minus pedestal value) was observed, with minimum and maximum values of approximately 25 mV and 50 mV. These values correspond to 0.31 $\pm$ 0.02 and 0.63 $\pm$ 0.02 photoelectron given that the peak amplitude average of a single photoelectron signal was measured at 79.5 $\pm$ 0.7 mV, as presented in figure~\ref{spe_spec}. Those values correspond to an efficiency of (99.4 $\pm$ 0.2)\% and (91.3 $\pm$ 1.5)\% in detecting SPE signals, respectively.
As a result, some channels were less sensitive to SPE signals than others, negatively affecting the reconstruction of low-energy events.

Making use of the threshold scan procedure, it was possible to measure the pedestal of each channel with great accuracy and precision. This measure could then be used to choose a single effective threshold value for all the channels.
Figure~\ref{fig:ThScan1} (right) shows the threshold scan curves after centering them in relation to their peak values (pedestal). As can be seen, an effective threshold of 25 mV could be applied to all the detector channels, equivalent to approximately 0.31 $\pm$ 0.02 photoelectron as before but for all the considered channels.

\begin{figure}[ht]
\centering
\includegraphics[width=0.49\linewidth]{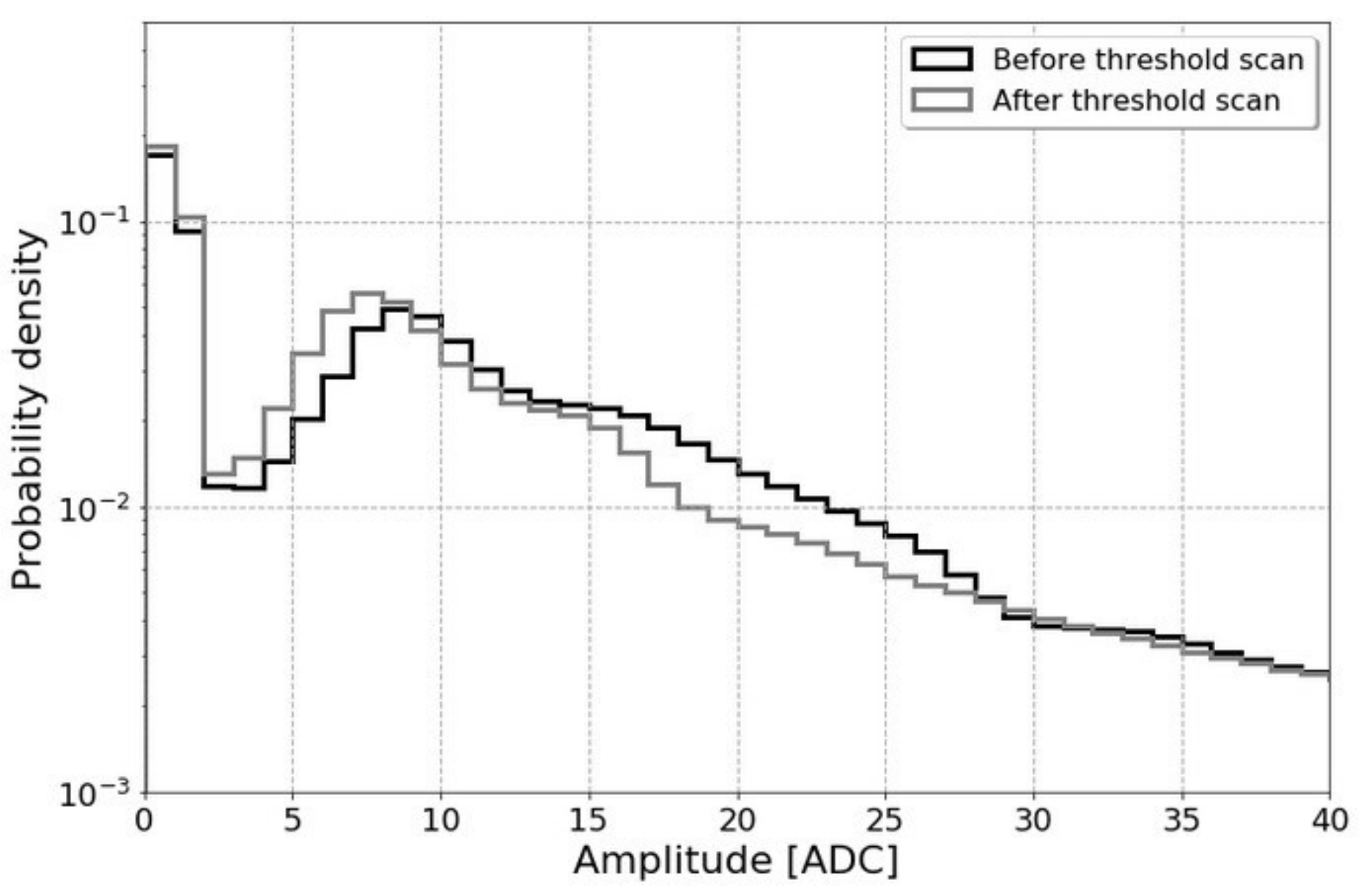}
\includegraphics[width=0.49\linewidth]{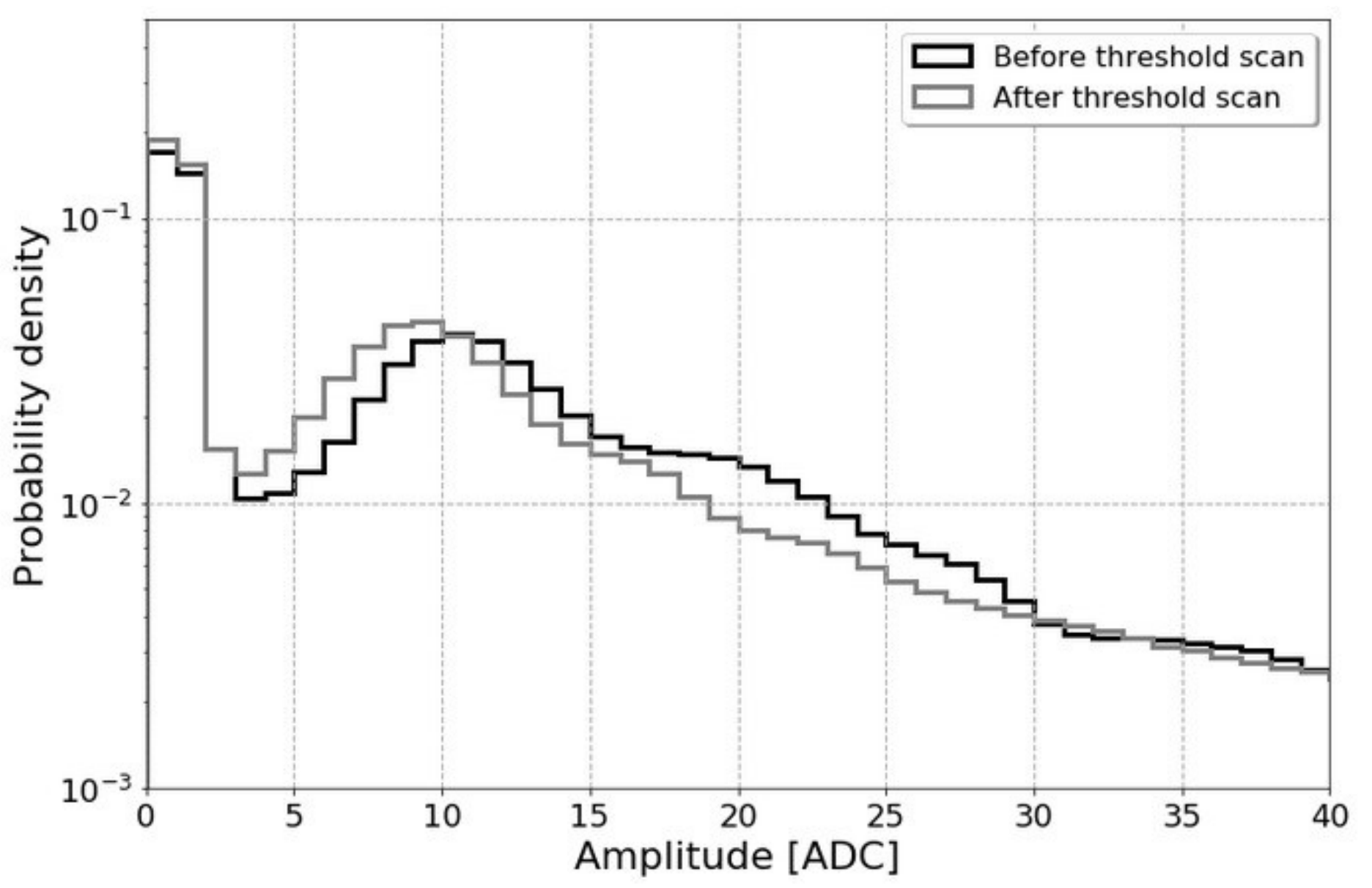}
\caption{Energy spectrum (in ADC units) for two of the TARGET detector channels before and after configuring their thresholds with the threshold scan procedure.}
\label{fig:DistEne1}
\end{figure}

Figure~\ref{fig:DistEne1} allows to analyze the impact of setting all the effective thresholds to 25 mV.
It shows the energy distribution (in ADC counts) for two channels of the TARGET detector where it is possible to observe that more signals with amplitude in the SPE region, below 10 ADC counts, were detected due to the increased efficiency in detecting SPE events (the $\nu$-Angra ADC resolution is 9.8~mV/count).

\begin{figure}[ht]
\centering
\includegraphics[width=0.6\linewidth]{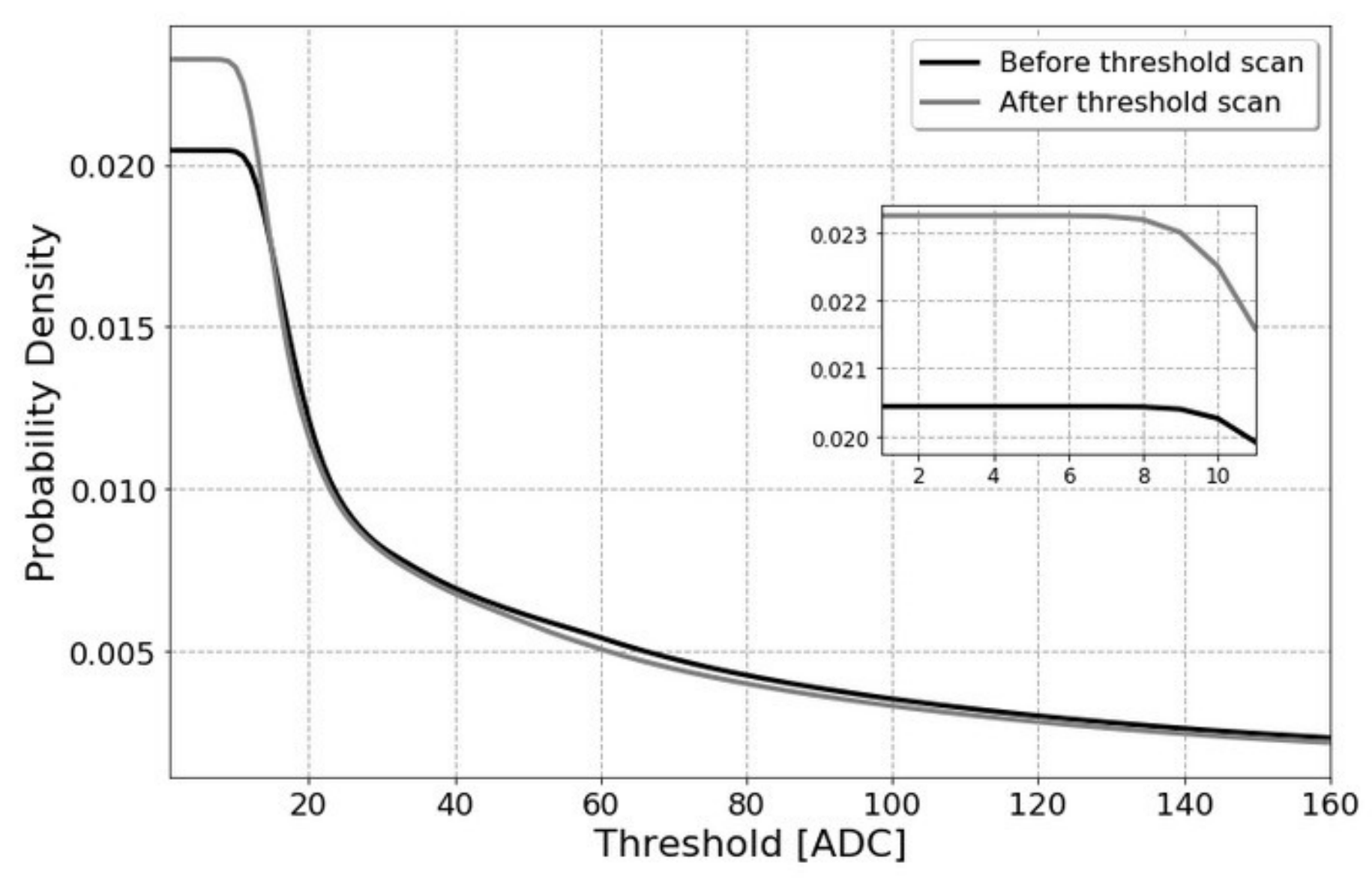}
\caption{Measurement of the probability density function of occurring an event with energy higher than a given threshold, for different threshold values, considering 60 hours of data acquisition. Only events which fired five or more channels were considered.}
\label{fig:IntEne}
\end{figure}

A measurement of the probability density function of occurring an event with energy higher than a given threshold, taken for different threshold values, is shown in figure~\ref{fig:IntEne}.
Two situations were considered: before and after using the threshold scan procedure to set the thresholds.
Only events which fired five or more channels were considered. Such coincidence requirement is used by the $\nu$-Angra experiment to reject events that might occur due to dark current pulses.
As can be seen, setting the thresholds of the front-end electronics discriminators based on the threshold scan procedure increases the number of low energy events around the SPE amplitude region by approximately 28\%.

The results presented in this section validate the proposed threshold configuration procedure showing that it is able to increase the probability of detecting low energy events, in the region of interest for antineutrino detection, providing a way to improve the overall performance of the $\nu$-Angra detector.

\section{Conclusions}
\label{sec:conclusion}

The main objective of this work was to describe the development of an embedded system aimed at the control and monitoring of fundamental parameters of the antineutrino detector of the \mbox{$\nu$-Angra} experiment. The results obtained during the development and commissioning phases and the importance of the developed system in the context of the application were presented.
The calibration and configuration of the detector's thresholds were treated as a central issue in this article since the events of interest for the experiment are at low energies, making it important to maximize the experiment's efficiency in detecting SPE signals.
This embedded system has been in use at the nuclear reactor site since the beginning of 2019 and all the detector channels have been configured using the threshold scan procedure.
It was able to set all the thresholds of the inner detector channels to 25.0 $\pm$ 1.2 mV, about three times less than the average peak amplitude of a SPE signal, measured to be at 79.5 $\pm$ 0.7 mV with a standard deviation of 21.7 $\pm$ 0.6 mV, making the detector's channels to reach an efficiency of (99.4 $\pm$ 0.2)\% in detecting signals generated by SPE events.

\acknowledgments
The authors would like to thank the Ministry of Education, the Ministry of Science, Technology and Innovation (MCTI) and the Funding Authority for Studies and Projects (FINEP) for supporting the Neutrinos Angra experiment whose initiative took place in 2007. We are also thankful to the Foundation for Research Support of
the State of Rio de Janeiro (FAPERJ) and the Foundation for Research Support of
the State of Minas Gerais (FAPEMIG) which, in 2011, approved the project TEC-APQ-02287/11 financing part of what was implemented at the Federal University of Juiz de Fora (UFJF). Finally, we thank the UFJF University for the excellent structure and support provided for the execution of this project.

\end{document}